\documentclass[11pt]{article}
\usepackage{graphicx}
\usepackage{amsmath}
\def\D{\Delta}
\def\d{\delta}

\def\S{\Sigma}
\def\G{\Gamma}
\def\g{\gamma}
\def\e{\epsilon}
\def\s{\sigma}
\def\S{\Sigma}
\def\o{\omega}

\def\a{\alpha}
\def\b{\beta}

\def\m{\mu}
\def\n{\nu}
\def\r{\rho}
\def\s{\sigma}
\def\p{\pi}

\def\f{\phi}
\def\F{\Phi}
\def\th{\theta}

\def\t{\tau}
\def\e{\epsilon}

\def\vf{\varphi}

\def\ch{{\cal H}}
\def\cd{{\cal D}}
\def\cg{{\cal G}}
\def\cl{{\cal L}}
\def\cs{{\cal S}}
\def\cr{{\cal R}}
\def\cv{{\cal V}}

\def\ha{\frac{1}{2}}
\def\pa{\partial}

\def\R{{\bf R}}
\def\C{{\bf C}}
\def\Z{{\bf Z}}

\def\bv{{\big |}}
\def\Bv{{\Big |}}

\newcommand{\be}{\begin{equation}}
\newcommand{\ee}{\end{equation}}
\newcommand{\bea}{\begin{eqnarray}}
\newcommand{\eea}{\end{eqnarray}}

\begin{document}

\begin{center}
\bf{\Large Generalization of Bohmian Mechanics and Quantum Gravity Effective Action}
\end{center}

\bigskip
\begin{center}
{\bf\large Aleksandar Mikovi\'c} 
\\
Departamento de Inform\'atica e Sistemas de Informac\~ao \\
 Lus\'ofona University and COPELABS\\
Av. do Campo Grande, 376, 1749-024 Lisboa, Portugal\\
and\\
Mathematical Physics Group, Instituto Superior T\'ecnico\\
Av. Rovisco Pais, 1049-001 Lisboa, Portugal
\end{center}

\bigskip
\bigskip
\begin{abstract} 
We generalize the de Broglie-Bohm (dBB) formulation of quantum mechanics to the case of quantum gravity (QG) by using the effective action for a QG theory. This is done by replacing the dBB equations of motion with the effective action equations of motion, which is beneficial even in the non-gravitational case, since in this way one avoids the violations of the Heisenberg uncertainity relations and the absence of the classical trajectories for stationary bound states. Another advantage of the effective action formalism is that one can obtain the field configurations in the case of a quantum field theory (QFT). The proposed QG generalization is natural for Bohmiam mechanics because a dBB wavefunction is really a wavefunction of the Universe and in order to define the effective action for an arbitrary initial state one needs a QG path integral.  The QG effective action can be constructed by using the piecewise flat quantum gravity (PFQG) theory and the PFQG effective action can be approximated by the QFT effective action for General Relativity coupled to matter, with a cutoff determined by the average edge length of the spacetime triangulation. One can then calculate the corresponding field configurations and from these field configurations one can obtain the trajectories for the corresponding elementary particles.
\end{abstract}

\newpage
\section{Introduction}
The standard interpretation of quantum mechanics (QM), see \cite{sqm}, is problematic for quantum cosmology, because it requires an observer outside of the system under consideration, while in a quantum gravity (QG) theory the system is the whole universe. However, in the de Broglie-Bohm (dBB) formulation of QM, see \cite{dbb,val}, there is no need for an observer, since the particles exist independently from the wavefunction, while in the standard QM the particle positions emerge only in a measurement process performed by an observer. Another advantage of the dBB QM is that the dBB wavefunction is really a wavefunction of the universe (WFU), so that the dBB formulation can be naturally used in a QG theory.

One can generalize the dBB QM for the case of gravitational minisuperspace models, see \cite{val}, but the case of full General Relativity (GR) has not been considered so far, because a QG theory with suitable properties has been lacking. There are several well-devoleped frameworks for a QG theory, most notably the string theory \cite{gsw,pol}, Loop Quantum Gravity \cite{ro,pe} and asymptotically safe QG \cite{asqg}, but the framework which is the most suitable for formulating a generalization of the dBB quantum mechanics is the piecewise flat quantum gravity (PFQG) \cite{MVb,M4,M5}. This is because the PFQG theory is based on a path-integral quantization of GR coupled to matter, and given that the PFQG path integral is finite \cite{M4}, one can determine the time evolution of the wavefunction of the Universe (WFU) and the corresponding effective action \cite{M5}. Since an effective action gives the quantum trajectories, the PFQG theory has the necessary ingredients to be a QG generalization of the dBB formulation.

Beside the problem of how to include gravity, one of the problems of the usal dBB QM is that it is not easy to generalize it to the case of a quantum field theory (QFT). A natural way would be to use the Schrodinger representation of a QFT, so that the basic elements would be a field configuration $\vf(\vec x,t)$ on the spacetime and a wavefunctional $\Psi[\vf(\vec x),t]$. The problem with this approach is that an interacting QFT in the Schrodinger representation is poorly understood and also it is not clear how to associate the particle trajectories to a field configuration $\vf(\vec x,t)$. That is why all the existing dBB generalizations are based on using the configuration space analog of the Fock space \cite{dbbqft}, which is an infinite direct sum of the Hilbert spaces for constant number of particles. However, then one has the problem of how to account for the creation and annihilation of particles, which requires the introduction of additional laws of motion. Namely, in each Hilbert space with a fixed number of particles one has the standard dBB equations of motion (EOM), but in order to describe the transitions between the different Hilbert spaces  one needs the additional laws of motion, and there is no a simple or a unique way to do this.

Beside a QG generalization of dBB QM we argue that the dBB EOM should be replaced by the effective action EOM. Note that Bohm formulated the dBB EOM in the second-order form, which were given as the Newton equations where the classical potential was replaced by the quantum potential \cite{bohm}. The inital conditions were given by the initial positions and the velocities (momenta), but this approach is not consistent, since the initial momenta cannot take arbitrary values in the dBB formulation. The reason for this is that the Bohm EOM are a consequence of the first-order dBB EOM.

However, if one instaed of the Bohm EOM uses the effective action EOM, one can have independent coordinates and momenta, and the inital-value problem is consistent. Another advantage of the effective action EOM is that one can explicitely realize the Heisenberg uncertainity relations (HUR), which are violated if one uses the dBB phase-space probability distribution. Also, it is easy to relate the quantum trajectories with the classical ones when $\hbar\to 0$ in the EA case since the effective action (EA) is given by the classical action plus the $\hbar$ corrections, so that one does not have a strange situation that the dBB trajectories for stationary bound states are static. In addition, it is easier to construct the QFT equations of motion (with a manifest Lorentz covariance) and these can be generalized to the case of full QG.

The main problem with the idea of replacing the dBB EOM with the EA EOM is in the fact that the QFT EA can be defined only for a single state, which is the  QFT vacuum state, so that it is not clear what to do for other states. This problem was solved in the context of a quantum gravity theory \cite{M5}, where it was shown that one can define an effective action associated to the time evolution of the WFU on a spacetime manifold $M_0 \sqcup(\S\times I)$, where $M_0$ is an arbitrary 4-manifold whose boundary is a 3-manifold $\S$, $I$ is a time interval and the initial state is given by the Hartle-Hawking state for the vacuum manifold $M_0$. This was done by using the path integral for the PFQG theory \cite{MVb,M4}, which has a finite path integral, so that the WFU and the effective action are well defined. Hence in this paper we will use these results in order to construct a dBB formulation of a QG theory.

In section 2 we review the standard QM formalism and the dBB reformulation. We also point out the problem of the Heisenberg uncertainity relations  violation in the dBB case and explain the problem of the classical limit of the dBB trajectories for stationary bound states. In section 3 we review the Schrodinger representation of a QFT, which would be a natural starting point for a dBB formulation. However, since it is difficult to work in the Schrodinger representation of a QFT, one passes to the Fock space representation. In the Fock space representation one can define the dBB particle trajectories in the fixed-number particle subspaces, but then it is not clear how to describe the particle trajectories when the particle number changes. We then propose to use the effective action EOM for field configurations, instead of the dBB particle trajectories, since the QFT effective action is well-defined perturbatively for a renormalisable QFT. However, the problem with this proposal is that the QFT effective action is only defined for the vacuum state, so that one does not know how to define the field  configurations for other states. We show how to solve this problem in section 4 by using the path integral for the PFQG theory. We also discuss the relationship
between the effective actions for the manifolds $M_0 \sqcup(\S\times I)$, $\S\times I$ and $M_0 \sqcup(\S\times I)\sqcup M_0$.

In section 5 we show how the PFQG effective action, which is defined on a piecewise linear manifold $T(M_0 \sqcup(\S\times I))$, where $T$ is a triangulation, can be approximated by a QFT effective action for GR coupled to matter fields on the smooth spacetime $\S\times I$ and how to compute the WFU correction. Consequently one can determine the EA field configurations and one can also determine the QFT wavefunctional that corresponds to the chosen WFU. Furthermore, given an EA field configuration, one can determine the corresponding particle trajectories. In section 7 we explain how to include the fermionic fields in this framework. In section 8 we present our conclusions. In appendix A we demonstrate the failure of the dBB phase-space probability distribution to give the same result as the QM expectation value for the square of the linear momentum, while in the appendix B we give a derivation of the dBB trajectory for a Hydrogen-atom bound state of constant energy and non-zero angular momentum. In appendix C we explain the relationship between the PFQG path integral and the QFT effective action and in the appendix D we explain how to calculate the corrections to the QFT effective action due to a non-trivial WFU.

\section{de Broglie-Bohm quantum mechanics}

The standard QM formulation is based on the classical phase-space of the physical system. Let us consider the phase space $\R^{2n}$ so that the particle positions, or the generalized coordinates, are given by a vector $q\in\R^n$, while the corresponding canonicaly conjugate momenta are given by a vector  $p\in\R^n$. The classical dynamics is determined by the Hamiltonian $H(p,q)$ and the corresponding EOM come from the Lagrangian $L = p\dot q -H$. 

The quantization is a map from the functions $f(p,q)$ on the phase space to a set of linear hermitian operators $\hat f$ acting on a Hilbert space such that 
\be \{f_1,f_2\}_{PB} = f_3 \to  [\hat f_1, \hat f_2] = i\hbar \hat f_3 \,,\ee
where the Poisson bracket is defined as
\be \{f_1,f_2\}_{PB} \equiv {\pa f_1\over\pa q}{\pa f_2\over\pa p} - {\pa f_1\over\pa p}{\pa f_2\over\pa q} \,.\ee

In the case of the phase-space coordinates we obtain the Heisenberg algebra 
\be [\hat q_k , \hat p_l] = i\hbar \d_{kl} \,,\quad [\hat p_k, \hat p_l] = 0\,,\quad [\hat q_k, \hat q_l] =0 \,, \label{heis}\ee
which has the Schrodinger representation on the Hilbert space $L^2(\R^n)$ given by
\be \hat p_k \Psi(q) = -i\hbar{\pa\Psi(q)\over\pa q_k} \,,\quad \hat q_k \Psi(q) = q_k\Psi(q) \,.\label{srep}\ee
The scalar product is defined by
\be \langle\Psi_1|\Psi_2\rangle = \int_{\R^n} \Psi_1^*(q)\Psi_2 (q) \,d^n q \,.\ee

The state of the sistem can be then represented by a wavefunction
$\Psi(q, t) \in L^2(\R^n)$,  whose time evolution is determined by the Schrodinger equation
\be i\hbar{\pa\Psi\over \pa t} = \hat H \Psi(q,t) \,,\ee
where $\hat H$ is a hermitian operator that corresponds to the classical hamiltonian $H(p,q)$ in the Schrodinger representation (\ref{srep}).

In order to connect the standard QM formalism with experiments, we need to introduce the measurement postulate. In standard QM, the unitary time evolution of a wavefunction is interrupted by an act of measurement, when an observer registers a value $\a$ of an observable $A$ of the system. This is described by the change of the state of the system $|\psi(t)\rangle$ at the moment $t=t_0$ of the measurement to the eigenstate        
$|\a\rangle$ of the operator $\hat A$. This situation is also described as the collapse of the wavefunction, because
\be \lim_{t\to t_0^+}\Psi(q,t) = \Psi_\a (q) \,,\ee
while $\Psi(q,t_0) \ne \Psi_\a (q)$, where $\Psi_\a (q) = \langle q|\a\rangle$.

The probability (or a density of probability if the spectrum of $\hat A$ is continious) of registering the value $\a$ is given by the Born rule
\be P(\a,t) = |\langle \a|\Psi(t)\rangle|^2 \,. \label{br} \ee

The formula (\ref{br}) is valid for the $n=1$ case, while for the $n>1$ case one has
\be P(\a,t) =\sum_{\a_2,...,\a_n} |\langle \a,\a_2,...,\a_n|\Psi(t)\rangle|^2 \,, \ee
see the formula (\ref{bpd}).

Hence the basic elements of the standard QM are the state $|\Psi(t)\rangle$, i.e. the wavefunction $\Psi(q,t)$, and the set of hermitian operators $\hat A_k$ that correspond to the classical observables $A_k (p,q)$ of the system\footnote{There are also observables which are not phase-space functions, like spin, but since we are focused on trajectories in space, these will not be essential for our purposes.}. The classical phase-space trajectories $(q(t), p(t))$ are not physical in the standard QM, because of the Heisenberg uncertainity principle. A configuration space trajectory can arise in standard QM as an extrapolation of a series of $q$-measurements, so that
\be \{ q(t)\, |\, t\in [a,b]\} \sim (q(t_1), q(t_2), \cdots, q(t_N)) \,,\ee
where $a\le t_1 < t_2 < \cdots < t_N \le b$. Hence the position of a particle in standard QM is not defined between the measurements.

The dBB QM resolves this conceptual problem by introducing a particle trajectory as a solution of the equation
\be p  = {\pa S\over \pa q} \,, \label{dbbem}\ee
where $\Psi(q,t) = R(q,t) \exp(i S(q,t)/\hbar)$. The equation (\ref{dbbem}) is a first-order differential equation
\be f(q, \dot q) = \hbar\,{\pa\over \pa q} Im\left( \log\Psi(q,t)\right)\,,  \ee
where $f(q,\dot q) = {\pa L\over \pa \dot q}$ and $L$ is the classical Lagrangian of the system. Hence the ontology of the dBB QM is a trajectory $q(t)$, which is determined by the dBB EOM (\ref{dbbem}), and a wavefunction $\Psi(q,t)$, determined by the Schrodinger equation. The probability of the system of having a coordinate value $q$ is given by the probability density distribution
\be \rho (q,t) = |\Psi(q,t)|^2 \,.\ee

Hence there is no a wavefunction collapse in dBB QM, since if we know that a particle is at a position $q_0$ at a moment $t_0$, this does not lead to the wavefunction collapse 
\be \lim_{t\to t_0^+}\Psi(q,t) = \d(q -q_0)\,,\ee
but $\Psi(q,t)$ stays continious at $t = t_0$, i.e.
\be \lim_{t\to t_0^-}\Psi(q,t) = \lim_{t\to t_0^+}\Psi(q,t)  \,.\ee

In the case of a variable $A(p,q)$, the probability of observing a value $\a$ is given by the probability distribution
\be \rho_A (\vec\a,t) = |\F(\vec\a,t)|^2 = |\langle \vec\a|\Psi(t)\rangle|^2 \,,\label{bpd}\ee
where $\vec\a =(\a,\a_2,...,\a_n)$ are the eigenvalues of a commuting set of linearly independent operators $\{\hat A, \hat A_2 , ... , \hat A_n\}$ such that $\hat A$ corresponds to the classical variable $A(p,q)$.


However, the momentum variables $p_k$ are special, because they are canonically conjugate to $q_k$. Their dBB trajectories are not independent from the $q_k$ trajectories, since $p_k = \pa S / \pa q_k$, so that the probability distribution of observing the particles at positions $q_k$ with momenta $p_k$ is given by
\be \r (p,q,t) = |\Psi(q,t)|^2 \, \prod_{k=1}^n\d\left(p_k - {\pa S\over \pa q_k}\right) \,. \label{pqbd}\ee

The problem with the distribution function (\ref{pqbd}) is that it may violate the Heisenberg uncertainity relations (HUR)
\be \D_\r p_k \,\D_\r q_k \ge \frac{\hbar}{2}\,,\quad k=1,2,...,n \,,\label{rhur} \ee
since
\be \langle p_k^2 \rangle_\r = \int d^n q\int d^n p \, p_k^2\, \r(p,q) = \int d^n q\, |\Psi(q,t)|^2 \,\left({\pa S\over\pa q_k}\right)^2 \,, \ee
is not the same as
\be \langle \hat p_k^2\rangle = \langle\Psi|\hat p_k^2 |\Psi\rangle = \int d^n q \,\Psi^*(q,t)\,(-\hbar^2)\,{\pa^2 \Psi(q,t) \over\pa q_k^2} \,,\ee
see the appendix A.

Consequently $\D_\r p_k \ne \D p_k$, where $(\D X)^2 = \langle X^2\rangle - \langle X \rangle^2$. Note that $\D_\r q_k = \D q_k$ and $\langle p_k\rangle_\r = \langle \hat p_k\rangle$, but $\langle p_k^2 \rangle_\r \ne \langle \hat p^2_k \rangle$. These relations open a possibility for a violation of the Heisenberg uncertainity relations (\ref{rhur}), which happens in the case of a stationary bound state, since then $\D_\r p = 0$, see the appendix A.

However, it is easy to see that $\D_\r p_k = \D p_k$ and $\D_\r q_k = \D q_k$ for
\be \r (p,q,t) =  |\Psi(q,t)|^2 |\F(p,t)|^2 \,,\label{hurd}\ee
where $\F(p,t)$ is the Fourier transform of $\Psi(q,t)$. In this case the HUR hold, and the distribution (\ref{hurd}) would be valid if the $p_k$ variables were dynamically independent from the $q_k$ variables, which happens in the case  of the EA equations of motion. We will explain this in the next section.

Note that the only way to obtain a complete dBB dynamics is to take the wavefunction to be a WFU. Then the wavefunction of a subsytem is defined as the conditional wavefunction, see \cite{dgz}. Let $q=(q_1,q_2)$ and $p=(p_1,p_2)$ such that $q_1$ are the coordinates of our system, while $q_2$ are the coordinates of the rest of the Universe. Then from the dBB EOM we have
\be p_1 = {\pa S\over \pa q_1} \,,\quad  p_2 = {\pa S\over \pa q_2} \,,\ee
so that we can obtain the dBB trajectories $q_k = f_k(t)$, $k=1,2$. We can then define the conditional wavefunction for our subsystem as
\be\tilde\psi_1(q_1, t) \equiv \Psi(q,t){\Big |}_{q_2 = f_2(t)} = \Psi(q_1, f_2(t), t)\,. \ee
Then the conditional probability distribution satisfies
\be \r(q_1,f_2(t) , t) = |\tilde\psi_1 (q_1,t)|^2 = \tilde\r_1(q_1,t) \,.\ee

However, the wavefunction $\tilde\psi_1(q_1,t)$ usually does not satisfy the Schrodinger equation for the subsystem, which is given by
\be i\hbar\pa_t\psi_1 = \hat H_1\psi_1 \,,\ee 
where $H_1(p_1,q_1)$ is the Hamiltonian for the subsystem. Let us assume that the total Hamiltonian has the form
\be  H =  H_1(p_1,q_1) +  H_2(p_2,q_2) +  H_{12}(p,q) \,,\ee
so that when the interaction between the system and the rest of the Universe is small, i.e. when $|\langle \hat H_k \rangle| \gg |\langle \hat H_{12} \rangle|$, $k=1,2$, during some time interval $\D t$, then $\Psi \approx \psi_1\psi_2$ and it can be shown that $\tilde\psi_1$ satisfies
\be i\hbar\pa_t \tilde\psi_1 \approx ( \hat H_1 + \D \hat H_1(t)) \tilde \psi_1 \,, \ee
see \cite{dgz}. The extra term $\D\hat H_1(t)$ then makes it possible for the conditional wavefunction to undergo a wavefunction collapse, while the WFU never collapses. When $\langle \D\hat H_1\rangle$ is negligible with respect to $\langle \hat H_1\rangle$ during a time interval $\D t' < \D t$, then the conditional wavefuction will obey the Schrodinger equation for the subsystem.

However, there is a feature of dBB QM which is problematic. Namely, the particles in a stationary bound state 
\be \Psi(q,t) = R(q) e^{-iEt/\hbar} \,,\ee
do not move, since $p_k = m_k\dot q_k$ (or more generally $p_k \approx m_k \dot q_k$ for $\dot q_k \to 0$) so that
\be m_k \dot q_k = {\pa\over \pa q_k} \left( -Et \right) = 0 \,\,\Rightarrow \,\, \dot q_k = 0 \,\,\Rightarrow \,\, q_k = \textrm{const.}\quad.\ee

Although it is difficult to imagine that an electron sits still inside a Hydrogen atom, one could say that this is not a problem, since the quantum trajectories are radically diferent from the classical ones. Even when a bounded electron can have a nontrivial dBB trajectory, which happens for the excited states in the energy spectrum with a non-zero angular momentum
\be \Psi_{n,l,m}(r,\th,\f,t) = R_n(r) P_l(\cos\th)e^{im\f}e^{-iE_n t /\hbar}\,,\quad m\ne 0\,,\ee
where $(r,\th,\f)$ are the spherical coordinates, $n \ge 2$, $l = 0, 1, ..., n-1$ and $|m| \le l$. In this case one obtains circular trajectories
\be r = const. \,,\quad \th = \pi/2\,,\quad \f = \o\, t + \f_0 \,,\label{ctdbb} \ee
where $\o = m\hbar / m_e r^2$, see the appendix B.

These type of trajectories for the Hydrogen bound states  are difficult to understand from the point of view of the classical limit, since for large $n$ one has semi-classical states, and one expects to see an almost classical trajectory, which is typically an ellipse. Another problem is that for a classical circular trajectory in a $1/r$ potential, it is easy to show that
\be \o \propto r^{-3/2}\,,\label{clo}\ee
while for the dBB trajectory (\ref{ctdbb}) we have
\be \o = {m\hbar \over m_e r^2} \propto r^{-2}\,.\label{dbbo}\ee

Although one can still say that the electrons in an atom do not have to have classical trajectories, the problem of the classical limit persits, since one can apply the same reasoning to the bound state system of the Sun and the Earth. Then it is not clear how to obtain a Kepler orbit from a dBB trajectory, since in the case of the Earth orbit we observe (approximately) the relationship (\ref{clo}), and not the relationship (\ref{dbbo}).

\section{de Broglie-Bohm QFT}

Before we explain the PLQG generalization of the Bohmian mechanics, we need to explain the dBB formulation of a QFT. Consider a field theory
\be S = \int_a^b dt\int_\S d^3x \left( \ha\dot\vf^2 - \ha \nabla\vf^2 - V(\vf) \right)\,, \ee
where the field $\vf(\vec x,t)$ is a Lorentz group scalar and the metric on $M =\S\times\R$ is taken to be flat. The potential $V(\vf)$ is assumed to be a polynomial function. The cases of a Dirac field and of a vector field can be treated in a similar way.

The Hamiltonian formulation is given by
\be S = \int_a^b dt\int_\S d^3x \left(\p\dot\vf - \ch(\vec x,t)\right)\,,\label{sfa}\ee
where $\p(\vec x,t)$ is the cannonically conjugate momentum to $\vf$ and
\be H = \int_\S d^3x \,\ch(\vec x,t) = \int_\S d^3x \left(\ha\p^2 + \ha(\nabla\vf)^2 + V(\vf) \right) \,,\label{qfth}\ee
is the Hamiltonian.

We can quantize the field theory (\ref{sfa}) by promoting the canonical pair $(\p(\vec x), \vf(\vec x))$ into operators acting on the vector space $\cv$ of functionals $\Psi[\vf(\vec x)]$ by using the Schrodinger representation
\be \hat\p(\vec x) \Psi[\vf] = - i\hbar {\d\Psi\over \d\vf(\vec x)} \,,\quad \hat\vf(\vec x)\Psi[\vf] = \vf(\vec x)\Psi[\vf] \,.\label{qftsr}\ee
For the sake of simplicity, we have supressed the dependence of the fields and the functionals on the parameter $t$.

A scalar product on $\cv$ can be defined via the path integral
\be\langle\Psi_1|\Psi_2\rangle = \int \cd\f \,\Psi_1^*[\vf]\, \Psi_2 [\vf] \,,\label{spi}\ee
and this is already a problem, since this path integral cannot be defined in general case. 

The wavefunctional $\Psi[\vf(x),t]$ obeys the Schrodinger equation
\be i\hbar{\pa\Psi\over\pa t} = \hat H \Psi[\vf,t] \,,\label{seft}\ee
where $\hat H$ is the operator associated to the Hamiltonian (\ref{qfth}) in the Schrodinger representation (\ref{qftsr}). Instead of the particle trajectories one now has the field configurations $\vf(\vec x,t)$ which are determined by the dBB EOM 
\be \dot\vf(\vec x,t) = {\d \cs\over \d\vf(\vec x,t)}\,,\label{dBBft}\ee
where $\Psi[\vf] = \cr[\vf]e^{i\cs[\vf]/\hbar}$ and we used $\p = \dot\vf$. The time evolution of an initial field configuration can be now determined.


However, it is still difficult to work in the Schrodinger representation of a QFT. Beside the problem of the scalar product, another problem is how to define the operator $\hat\p^2(\vec x)$ in the Hamiltonian operator, since one needs to regularize 
\be \hat\p(\vec x)\hat\p(\vec y)\Psi[\vf] = \Psi_1 [\vf] + \Psi_2[\vf]\d(\vec x-\vec y)
 + \Psi_3 [\vf]\nabla^2_x \d(\vec x-\vec y) + \cdots \,, \ee
when $\vec x\to \vec y$. In addition, a very little is known about the solutions of the functional Schrodinger equation (\ref{seft}). 

That is why the Fock space representation is used instead of the Schrodinger representation. The Fock space is defined by the vectors
\be |\Psi\rangle =\a|0\rangle +  \sum_{n \,\ge\, 1}^\infty\int_{\S^*} d^3k_1 \cdots \int_{\S^*} d^3k_n\,  c(\vec k_1,...,\vec k_n)\, \hat a^\dagger (\vec k_1)\cdots \hat a^\dagger (\vec k_n)|0\rangle \,,\label{fst}\ee
where $\S^*$ is the dual set of $\S$ for the Fourier transform\footnote{For example, when $\S =\R^3$, then $\S^* = \R^3$. However, if $\S$ is a 3-torus, then $\S^* = \Z^3$.}, while $\hat a^\dagger$ and $\hat a$ are the creation and the annihilation operators, defined by
\be  \hat a(\vec k) = (2\pi)^{-3/2}\int_\S d^3x \, e^{i\vec k \vec x}\left({i\hat\p(\vec x) + \o_k \,\hat\vf(\vec x)\over \sqrt{2 \hbar\o_k}} \right) \,,\ee
and
\be\hat a^\dagger (\vec k) =(2\pi)^{-3/2} \int_\S d^3x \, e^{-i\vec k \vec x} \left({-i\hat\p(\vec x) + \o_k \,\hat\vf(\vec x)\over \sqrt{2\hbar\o_k}} \right)\,.\ee
The frequency $\o_k = \sqrt{k^2 + m^2}$, where $k^2 = \vec k \cdot \vec k$ and $m^2 = V''(0)$. When $m=0$, $\o_0$ is given by an arbitrary non-zero constant.

The creation and annihilation operators obey the algebra
\be [\hat a(\vec k), \hat a (\vec q) ] = 0\,,\quad [\hat a(\vec k), \hat a^\dagger (\vec q) ] = \d(\vec k - \vec q)\,,\quad [\hat a^\dagger (\vec k), \hat a^\dagger (\vec q) ] = 0\,, \ee
while the perturbative vacuum state is defined as
\be \hat a(\vec k) |0\rangle = 0 \,,\quad\forall \,\vec k \in \S^*\,. \ee

The scalar product in the Fock space can be defined as
\be \langle\Psi_1|\Psi_2\rangle_F = \a_1^*\,\a_2 + \sum_{n\,\ge\,1}^\infty \int_{\S^*} d^3k_1 \cdots \int_{\S^*}  d^3k_n\,  c_1^*(\vec k_1,...,\vec k_n) \, c_2 (\vec k_1,...,\vec k_n) \,.\ee
This scalar product can be asilly evaluated, and represents a sum of the usual QM scalar products for $n$-particle states in the momentum-space representation. 


Given a Fock state (\ref{fst}), one can define the $n$-particle wavefunctions for $n=1,2,...$ as
\be\psi_n (\vec x_1, ..., \vec x_n, t) = \int_{\S^*} d^3k_1 \cdots \int_{\S^*} d^3k_n\, e^{i(\vec k_1 \vec x_1 + ... +\vec k_n \vec x_n)} c(\vec k_1, ...,\vec k_n,t) \,. \ee
The time dependence is determined from the Schrodinger equation
\be i\hbar{\pa|\Psi(t)\rangle\over\pa t} = \hat H_F |\Psi(t)\rangle \,,\ee
where
\be\hat H_F = \int_{\S^*}  d^3k \,\hbar\o_k \,\hat a^\dagger(\vec k) \hat a(\vec k) + \cdots \,.\ee
The $\cdots$ correspond to the $\int_\S d^3x \,[V(\vf)-m\vf^2/2]$ term in the Hamiltonian which is expressed via the creation and annihilation operators.

One can then define the usual dBB dynamics in each $n$-particle sector via
\be {\vec p}_k^{\,(n)} = m{d\vec x_k\over dt} = \hbar {\pa\over\pa \vec x_k}Im(\log\psi_n (\vec x_1, ..., \vec x_n, t) )\,,\quad n=1,2,... \,\,,\label{dbbf} \ee
see \cite{dbbqft}. However, the equations (\ref{dbbf}) are not sufficient for the determination of the particle trajectories, since in a QFT one can have transitions between the diferent $n$-particle spaces, so that one needs additional equations of motion to the dBB equations (\ref{dbbf}). However, it is not clear how to choose these extra equations in a simple or a unique manner, so that several proposals have been made, see \cite{dbbqft, bqft}.

The problem of how to define the particle trajectories in a QFT can be avoided by using the field configurations. However, it is difficult to work in the Schrodinger representation of a QFT, while it is not clear how to define a wavefunctional in the Fock representation. One can then stay in the Fock representation, but replace the dBB equations of motion (\ref{dBBft}) with the effective action equations of motion
\be {\d\G[\vf]\over\d\vf(\vec x,t)} = 0 \,,\label{qftem}\ee
where $\G[\vf]$ is the QFT effective action. For a renormalizable QFT, it is well known how to calculate the perturbative effective action, see \cite{qft}. In this way one can introduce the field configurations in a QFT, which evolve in time according to the quantum corrected classical EOM, since
\be \G[\vf] = S[\vf] + \hbar\G_1[\vf] + \hbar^2 \G_2[\vf] + \cdots \,,\label{sce}\ee
where $S[\vf]$ is the classical action (\ref{sfa}). Consequently a solution of the equation (\ref{qftem})  will be given by
\be \vf(x) = \vf_0(x) + \hbar\vf_1(x) + \hbar^2 \vf_2(x) + \cdots \,,\label{qps}\ee
where $\vf_0$ is a classical solution and $x=(\vec x,t)$.

The effective action can be also defined for a system with finitely many degrees of freedom (DOF), see \cite{cdmp} for the case of an anharmonic linear oscilator, so that in the QM case one can also substitute the dBB EOM by the effective action EOM 
\be {\d\G[q]\over\d q(t)} = 0 \,.\label{qpem}\ee

This modification of the dBB EOM resolves the problem of the initial conditions, because $\G$ is a functional of $q(t)$ and $\dot q(t)$ ($\G$ can also depend on the higher-order derivatives of $q(t)$), so that one can freely specify the initial coordinates and the velocities. This also implies that the coordinates and the canonically conjugate momenta are dynamically independent, so that one can use the distribution (\ref{hurd}) which guarantees the Heisenberg uncertainity relations. When $\hbar\to 0$, we have the particle analog of the semi-classical expansion (\ref{qps}), so that there will be a clear relationship between the quantum and the classical trajectories for the semi-classical states. 

However, the problem with the standard QFT effective action is that it is only defined for the QFT vacuum state $|\Psi_0 \rangle$, so that it is not clear how to define $\G$ for some arbitray state $|\Psi\rangle$. As shown in \cite{M5}, this can be done by using the path integral which includes the gravitational DOF. This approach requires a QG theory with a finite path integral such that the semi-classical expansion (\ref{sce}) is valid. The PFQG theory \cite{MVb, M4} is an example of such a theory, and in the next section we will explain how to construct the QFT effective action for an arbitrary inital state.

\section{PFQG and quantum mechanics}

The PFQG theory, see \cite{MVb, Mrev}, is based on the assumption that the short-distance structure of the spacetime is not a smooth four-dimensional manifold $M$ but a piecewise linear (PL) manifold $T(M)$, where $T$ is a triangulation of $M$. Hence a spacetime triangulation is a physical feature of the spacetime, and not just an auxiliary tool which serves to define the path integral. The gravitational DOF are the edge lengths, so that one can define a PL metric which is flat in each 4-simplex and the corresponding path integral is given by the Regge path integral (PI), but with a non-trivial PI measure. This non-trivial PI measure is responsible for the finiteness of the path integral and for the correct semi-classical expansion of the effective action.

The smooth spacetime $M$ is then an approximation which appears at the distances larger then the Planck length, and this happens when the triangulation has a large number of the edges $N$ and the edge lengths are small, of the order of $l_0/N$, where $l_0$ is a free length parameter. In this case the dominant physics is determined by the QFT for GR coupled to matter on $M$, with a cutoff proportional to the inverse average edge length in $T(M)$.

Given a PL manifold $T(M)$, let $\{L_\e \,|\,\e = 1,2,...,N\}$ be a set of the edge lengths, such that $L_\e \in \R_+$ for a spacelike edge, while $L_\e \in i\R_+$ for a timelike edge. The choice of the timelike and the spacelike edges can be determined by the embedding of $T(M)$ into a five-dimensional Minkowski space. If $M$ is non-compact, we will assign the non-zero edge lengths only to a compact subset $T(M_c)$ of $T(M)$, and $M_c$ can be choosen to be a $B_4$ or $B_3 \times [0,1]$ where $B_k$ is a $k$-dimensional ball. These choices are related with the topological restriction we impose on $M$, which is given by
\be M = M_0 \sqcup (\S\times [0,1])  \,,\label{eam}\ee
where $M_0$ is an arbitrary four-manifold, while $\S$ is a three-dimensional submanifold of $M_0$, see Fig. 1
\begin{figure}[htpb] 
\centering
\includegraphics[width=0.8\textwidth]{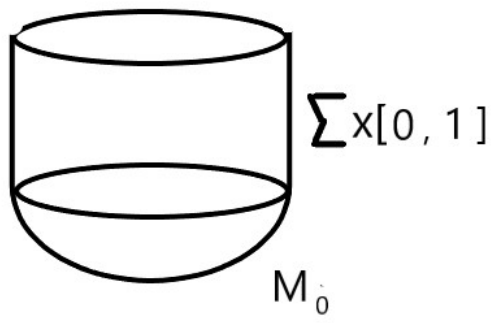}
\caption{The topology of a PFQG spacetime manifold}
\end{figure} 

Let $\F = \{\vf_v \,|\, v =1,2,...,n\}$ be a set of a matter field values at the vertices of $T(M)$. When $M$ is a non-compact manifold, we use the vertices of the compact subset $T(M_c)$. Then we will define the PFQG path integral as
\be Z_T(M) = \int_{D(T)} d^N L\, \m(L)\, e^{iS_R(L)/l_P^2} Z_m (L)\label{plqg} \ee
where $S_R(L)$ is the Regge action, $l_P = \sqrt{G_N\hbar}$ is the Planck length and $G_N$ is the Newton constant. The Regge action is the Einstein-Hilbert action $\int_M \sqrt{g}\, R(g)\,d^4x$ on $T(M)$ where a smooth metric $g(x)$ on $M$ is replaced by a PL metric on $T(M)$ defined by a set of flat metrics $g^{(\s)}(L)$, where $\s$ is a 4-simplex and $L$ are the edge-lengths of $\s$. 

The integration region $D(T)\subset \R_+^N$ is determined by the triangle inequalities for the spacelike triangles, while
\be Z_m(L) = \tilde Z_m (\tilde L)\bv_{\tilde L = w(L)} \,,\ee
is the matter path integral, where
\be \tilde Z_m (\tilde L) = \int_{D_m} \cd \F \, e^{- S_m(\tilde L,\F)/\hbar}\ee
is the Eucledean path integral for the matter fields and 
\be D_m = \R^{c_b n}\times \prod_{v=1}^n \cg_v(2c_f) \,,\label{mpir}\ee
where $c_b$ is the number of bosonic fields components and $\cg_v (d)$ is a Grassman algebra of dimension $2^{d}$ which is used to integrate $c_f$ complex fermionic components at a vertex $v$, see section 6, while
\be \cd\F = \prod_{v=1}^n d^{c_b}\vf_v  \, d^{c_f}\bar\th_v \, d^{c_f}\th_v \,. \ee

The action $S_m(L,\F)$ is the matter action $S_m[g(x),\vf (x)]$ on $T(M)$, where $\vf (x)$ is a collection of smooth matter fields on $M$, while $\tilde L = w(L)$ is a Wick rotation of a vector $L =(L_1,L_2,...,L_N)$, given by $\tilde L_\e = |L_\e|$, see \cite{M4}.

The PI measure $\m(L)$ should be chosen such that the path integral (\ref{plqg}) is finite and that it gives the correct semi-classical expansion of the effective action.  This can be achieved by using 
\be \m(L) = e^{-V_4 (L)/L_0^4}\prod_{\e=1}^N \left(1+ {|L_\e|^2 \over l_0^2} \right)^{-p} \,,\label{pim}\ee
with $p >$ 52.5 for the Standard Model, where $L_0$ and $l_0$ are free parameters of the theory \cite{M4}.

In order to construct an effective action for an arbitrary inital WFU, one has to use a time ordered triangulation of $\S\times [0,1]$. This can be achieved by embedding $\S\times [0,1]$ into $\R^5$ with a Minkowski metric such that $\S$ is embedded into a spacelike 4-plane of $\R^5$ while the interval $[0,1]$ is embedded in an interval $[t_i,t_f]$ of a timelike line of $\R^5$. We divide the interval $[t_i,t_f]$ into $n-1$ subintervals, and at each slice we introduce a triangulation of $\S$, $T_k (\S)$, $k=1,2,...,n$. We will choose the edges of a triangulation $T_k (\S)$ to be spacelike, while the edges that connect a pair $(T_k, T_{k+1})$ for $k=1,2,...,n-1$, can be choosen to be timelike.

We will call a time-ordered triangulation a temporal triangulation\footnote{A special case of the temporal triangulations, where all the spacelike lengths are equal and all the timelike lengths take only two values is called a causal triangulation \cite{cdt}.}. One can define a discrete time variable on a temporal triangulation as
\be t_k - t_i = \max \{L_\g\, | \,\pa\g_k = \{v\in T_1^{(0)} , v'\in T_k^{(0)} \}\}\,, \quad k=2,3,...,n\,,\label{tv}\ee
where $\g$ is a timelike line which connects the inital and the final triangulation and
\be L_\g = \sum_{\e\in\g} |L_\e| \,.\ee

One can further restrict a temporal triangulation such that all the spacelike triangulations $T_k(\S)$ are the same. We will call such a triangulation a Hamiltonian triangulation and in this case we can order the set of the vertices in each $T_k (\S)$ as $(v_1(k),v_2(k), ... , v_m(k))$ such that there is a timelike edge $L_\e(k,l)$ connecting $v_l(k)$ and $v_l(k+1)$. We can then introduce a discrete evolution parameter $t$ through a gauge fixing
\be  t_k -t_i = \sum_{k'=1}^{k-1}\max\{ |L_\e (k',l) | : l =1,2,...,m\} \,.\label{tgf}\ee

We can simplify the gauge fixing function by requiring $|L_\e(k',l)| = \D t_{k'}$ for $l =1,2,...,m$,
so that the gauge choice (\ref{tgf}) becomes
\be t_k - t_i = \sum_{k'=1}^{k-1} \D t_{k'}  \,.\label{htv}\ee

Given a temporal triangulation $T(\S\times[0,1])$ such that $T_1(\S) = T_k(\S)$, with the time variable given by (\ref{tv}), we can define the time evolution of a wavefunction as
\be\Psi(q,t_k) = Z_T(M_0\sqcup(\S\times [0,1])) = Z_T(M) \,,\label{qgte}\ee
where $q = (l,\vf)$, such that $l = \{L_\e \,|\,\e \in T^{(1)}_k(\S)\}$ and $\vf = \{\vf_v \,|\, v\in T^{(0)}_k (\S)\}$, see Fig. 2. 
\begin{figure}[htpb] 
\centering
\includegraphics[width=0.8\textwidth]{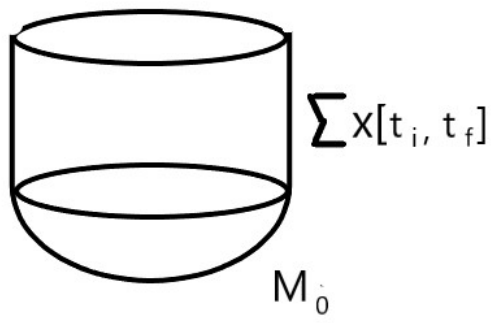}
\caption{PFQG spacetime manifold with a time variable interval.}
\end{figure} 
The manifold $M_0$ determines the inital wavefunction by the formula
\be \Psi_0 (q) = Z_{T}(M_0) \,,\label{hh}\ee
so that $\Psi_0(q)$ is the Hartle-Hawking (HH) wavefunction\footnote{More precisely, it is a Lorentzian generalization of the HH wavefunction, since in the path integral  (\ref{hh}) one uses the Lorentzian weight $e^{iS/\hbar}$ instead of the Euclidean weight $e^{-S/\hbar}$.}, see Fig. 3
\begin{figure}[htpb] 
\centering
\includegraphics[width=0.6\textwidth]{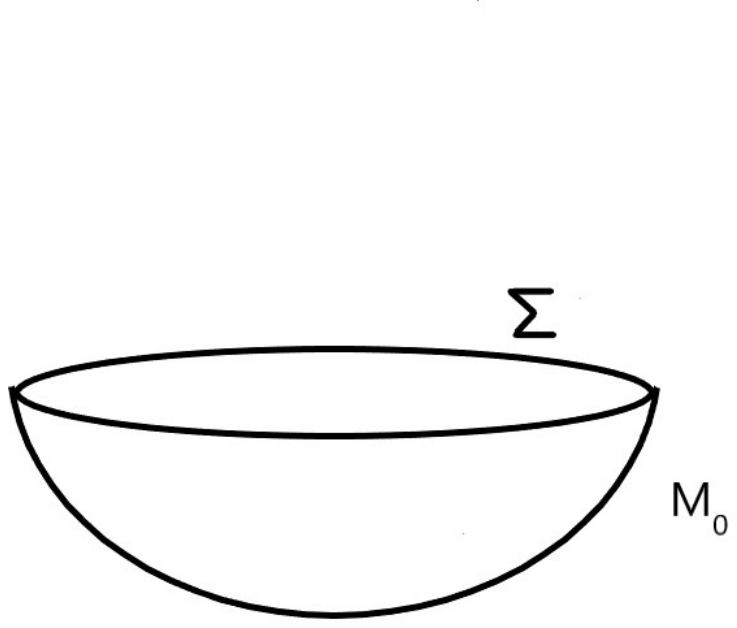}
\caption{Hartle-Hawking manifold}
\end{figure} 

For a sufficiently large $n$, we can consider $t_k$ as a continious variable $t\in[t_i,t_f]$, and we can write
\be \Psi(q,t) = \hat U_T(t) \Psi_0(q) \,,\ee
where $\hat U_T (t)$ is the QG analog of the QM evolution operator. Because $\hat U_T$ is defined via the path integral (\ref{qgte}), then
\be \hat U_T(t')\hat U_T (t) = \hat U_T (t'+t) \,.\ee
However, whether $\hat U_T$ is a unitary operator or not, this depends on the choice of the triangulation $T(\S\times[0,1])$. It is reasonable to expect that $\hat U_T$ will be a unitary operator for a Hamiltonian triangulation $T$.

Since we use the time variable (\ref{tv}) for $T(\S\times[0,1])$ such that the evolution parameter is in the interval $[t_i, t_f]$, then the choice of the manifold given by (\ref{eam}) can be represented as
\be M =  M_0 \sqcup (\S\times [t_i,t_f]) = M_0 \sqcup U \,,\label{team}\ee
where $U=\S\times [t_i,t_f]$, see Fig. 2.

Given the inital WFU (\ref{hh}) and the corresponding WFU at a time $t=t_f$, given by the path integral (\ref{qgte}), then the corresponding effective action $\G(L,\F)$ will give the quantum trajectories in $T(M)$. This EA is determined by the generating functional
\be Z_M (J) = \int \cd Q \,\m(L) \,e^{\frac{i}{\hbar}[S(Q)  + J Q] }\,,\ee
where $Q = (L, \F)$, $\cd Q = d^N L \cd \F$, $J =(J_L, iJ_\f)$, $J Q = J_{L} L + i J_{\F} \F$ and
\be  S(Q) = \frac{1}{G_N} S_R (L) + i\tilde S_m (L,\F) \,,\label{pica}\ee
where $\tilde S_m (L,\F) = S_m(\tilde L,\F)$ and $\tilde L = w(L)$. We also simplify the notation $Z_T(M,J)$ to $Z_M(J)$.

It will be useful to decompose the vectors $Q$ and $J$ as 
\be Q = (Q_-, q_-, Q_U, q_+) \,, \quad J = (J_-, j_-, J_U, j_+) \,, \ee
where the components refer to the values on the manifolds
\be \left( T(M_0)\setminus T_i(\S),\, T_i(\S), \, T(U)\setminus\{T_i(\S), \, T_f(\S)\}, \, T_f(\S)\right) \,,\ee
respectively. 

Since the Legandre transform of $W_M(J) = -i\hbar\log Z_M(J)$ gives the effective action, i.e.
\be \G_M(Q) = W_M(J) - JQ \,, \ee
where $Q = {\pa W_M \over \pa J}$, then we will have 
\be \G_M ( Q_-, q_-, Q  , q_+) = W  (J_- , j_-, J_U , j_+) -  J_- Q_- - j_-q_- -  J_U Q_U - j_+q_+ \,,\label{hhea}\ee
where 
\be Q_- = {\pa W_M\over \pa J_-} \,,\quad q_- = {\pa W_M\over \pa j_-}\,, \quad Q_U = {\pa W_M\over \pa J_U}\,,\quad q_+ = {\pa W_M\over \pa j_+}\,.\ee
The corresponding EOM are then given by
\be {\pa\G_M \over\pa Q_-} = 0\,,\quad {\pa\G_M\over\pa q_-} = 0 \,,\quad {\pa\G_M \over\pa Q_U} = 0 \,,\quad {\pa\G_M\over\pa q_+} = 0 \,.  \ee

These equations will determine the set of quantum configurations $Q$ on $T(M)$. Note that there may be more than one stationary point $Q_0$ of the function $\G_M (Q)$. 

In the case of a Hamiltonian triangulation of $U$ with a time variable (\ref{htv}), then for $\D t_k \to 0$, we will have
\be Q_U^0 \approx \{\{q_0(t)\,|\, t\in(t_i,t_f)\}, L(\D t) \} \,, \label{eaht}\ee
where $L(\D t)$ indicates the set of timelike edge lengths in $T(U)$. These edge lengths are not independent variables, but they are functions of $l(t_k)$ and $\D t_k$. Also note that $L_\e(\D t) = O(\D t_k)$.

Note that the trajectory $q_0(t)$,  $t\in [t_i,t_f]$, can be generated as a solution of the EOM for the EA
\be \G_H[q(t), \dot q(t), \ddot q(t), \cdots ] \approx \G (Q_-^0, q_-^0,Q_U , q_+^0)\,. \ee

The corresponding EOM 
\be{\d \G_H\over\d q(t)} = 0\,\Leftrightarrow\, {\d \G_H \over\d l_\e(t)} = 0\,,\, {\d \G_H\over\d \vf_v(t)} = 0 \,,\label{plhem}\ee
for all $\e\in T^{(1)}(\S)$ and all $v\in T^{(0)}(\S)$, will determine the quantum trajectories.

Note that there will be infinitely many quantum trajectories which have the classical inital data $(q_0,v_0)$ since a particular solution of the EOM (\ref{plhem}) will be determined by the inital conditions
\be q(0) = q_0 \,,\quad \dot q(0) = v_0 \,,\quad \ddot q(0) = a_0 \,,\,\,  \cdots \,. \label{siv}\ee 
However, only a subset of these trajectories will correspond to $q_0(t)$ trajectories. It may happen that there is only one $q_0(t)$ trajectory, which is always the case for the dBB EOM. However, in the PLQG case we are dealing with the discrete field configurations, so that the particle trajectories are not fundamental, and they will apear only in the smooth manifold approximation, which we will show in the next section.

Note that in the case of a QFT we have the following property
\be q_{EA}(t) = \langle\Psi_0| \hat q_{QFT}(t)  |\Psi_0\rangle  = {\pa \tilde W_U (\tilde J)\over\pa \tilde J(t)}\Bv_{ \tilde J=0}\,,\ee
where 
\be \hat q_{QFT} (t) = \hat U(t_f,t) \,\hat q\, \hat U(t,t_i) \,, \ee
and we have denoted $J_U$ as $\tilde J$, while  $\tilde W_U = -i\hbar\log \tilde Z_U$, see the appendix C. The generating functional $\tilde Z_U (\tilde J)$ is given by the generating functional $Z_{\bar M}(\bar J)$ for special values of the current vector $\bar J = (J_-, j_-, \tilde J, j_+, J_+) $ on the manifold
\be \bar M = M_0\sqcup U \sqcup M_0 \equiv M_- \sqcup U \sqcup M_+ \,,\ee
see Fig. 4, where the special values of the currents are given by
\be J_- = 0\,,\quad  j_- = 0\,,\quad \tilde J\ne 0\,,\quad j_+ = 0\,,\quad J_+ =0 \,. \ee

\begin{figure}[htpb] 
\centering
\includegraphics[width=0.6\textwidth]{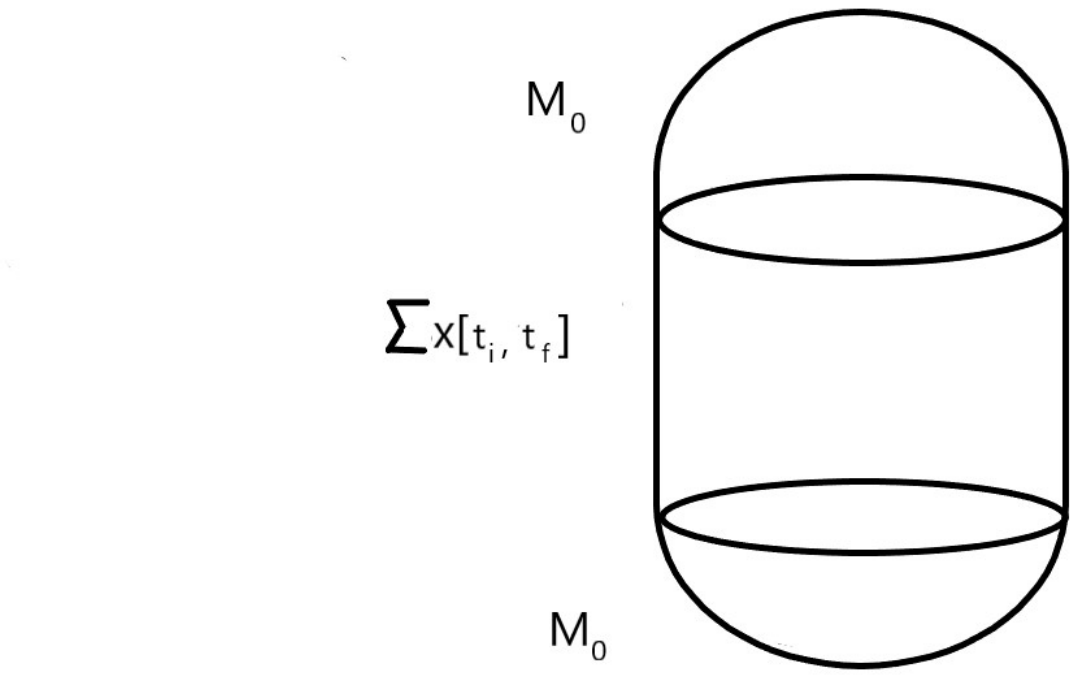}
\caption{QFT manifold}
\end{figure} 

The usual QFT effective action is a smooth approximation of the effective action $\G_U(\tilde Q)$, which is obtained  from $\tilde Z_U(\tilde J)$ for a trivial WFU, given by 
\be \Psi_0(q) = \d (q-q_0) = \prod_{\e\in T(\S)} \d(l_\e - l_\e^0)\prod_{v\in T(\S)}\d(\vf_v) \,,\ee 
where the vector $l_0$ correspond to a flat PL metric on $T(\S)$.

Although the effective action $\G_U(\tilde Q)$ will take into account the effects of a nontrivial WFU, it is not easy to work with it\footnote{For example, it is not clear  how to obtain a perturbative $\hbar$-expansion of $\tilde\G_U$.}, so that it is easier to use the effective action $\G_M$ and the related effective action $\G_U$ in order
to determine the QFT field configurations. We will explain this in the next section. 

Note that the values of $q_{EA}(t)$, as well as the values of $Q_M^0$, will be in general complex numbers, since the corresponding effective actions wil be complex-valued functions. This problem is resolved in a flat-metric QFT by using the Wick rotation and the Eucledean path integral, while in the PFQG case this problem can be resolved by using a real effective action, defined by
\be \G \to Re\,\G + Im\, \G \,,\ee
see \cite{MVb}.

\section{The EA trajectories in PFQG}

The QFT effective action has the perturbative expansion (\ref{sce}) which is the reason why there are quantum field configurations which are close to the classical ones. This also happens in the case of the PFQG effective action, so that there are quantum trajectories which are close to the classical ones. 

The effective action $\G(L,\F)$ for a PL manifold $T(M)$  obeys the EA equation
\be  e^{\frac{i}{\hbar}\Gamma(L,\F)} = \int_{D(L)} d^N  l \int_{D_m}  d^{cn} \vf \, \mu (L +l)\, e^{\frac{i}{\hbar}\left[S (L+l, \F + \vf) - \sum_{\epsilon} \G'_\epsilon (L,\F)l_\epsilon  -  \sum_{\p} {\Gamma}'_\p (L,\F) \vf_\p \right] } \,,\label{gmea}\ee
where $D(L)\subset \R^N$ is the region $D\subset\R_+^N$ translated by the vector $-L$, while $D_m$ is given by (\ref{mpir}). The action $S$ in (\ref{gmea}) is given by (\ref{pica}), while the manifold $M$ in (\ref{gmea}) can be an arbitrary 4-manifold. For the purposes of this paper we will restrict our choice to
$M = M_0 \sqcup (\S\times[t_i,t_f])$
such that the triangulation of $M_0$ can be arbitrary, while the triangulation of the manifold $U = \S\times[t_i,t_f]$ must be a temporal triangulation.

The equation (\ref{gmea}) is the PFQG generalization of the QFT effective action equation
\be e^{i\G[\vf]/\hbar} = \int \cd\f \exp\left[\frac{i}{\hbar}\left(S[\vf + \f] - \int_M {\d\G[\vf]\over \d\vf(x)} \f(x) \,d^4 x\right)\right] \,,\ee
where now $\vf$ is a collection of smooth fields on $M$, which can include the metric on $M$, see \cite{MVb}.

When $|L_\e| > l_P$ for all the edges $\e$ in $T(M)$ and $|\tilde\vf_\a(v)| < 1$, for all the vertices $v$ in $T(M)$ and $\a = 1,2,...,c$, where $\tilde\vf = \sqrt{G_N}\vf$, it can be shown that a solution of (\ref{gmea}) is given by a perturbative series
\be \G(L,\F) = S(L,\F) + \hbar\G_1(L,\F) + \hbar^2 \G_2(L,\F)   + \cdots \,,\label{plpe}\ee
while the functions $\G_k (L,\F)$, $k=1,2,...$, are uniquely determined by the classical action $S(L,\F)$ and the path-inegral measure $\m(L)$, see \cite{MVb}.

The smooth-manifold approximation is realized when $N\to\infty$ and $L_\e = O(l_0/N)$ for all $\e$ in $T(M)$, where $l_0$ is the parameter of the PI measure (\ref{pim}). In this case there is a smooth metric $g_{\m\n} (x)$ on $M$ such that 
\be g_{\m\n}(x) \approx g_{\m\n}^{(\s)}(L)\,,\quad x\in\s  \,,\ee
for any 4-simplex $\s$ in $T(M)$ and there are smooth fields $\vf_\a(x)$ on $M$, $\a = 1,2,...,c$, such that
\be \vf_\a (x) \approx \vf_\a (v)\,,\quad  x\in v^* \,,\ee
for any vertex $v$ in $T(M)$,  where $v^*$ is the dual cell for a vertex $v$.

Then the smooth-manifold approximation is valid, and the perturbative expansion (\ref{plpe}) gives
\be \G(L,\F) \approx \G_{M}[g,\vf] = S[g,\vf] + \hbar\G_{M, 1}[ g,\vf] + \hbar^2 \G_{M,2}[g,\vf]   + \cdots \,,\label{smpe}\ee
where the coefficients $\G_{M,k}[g,\vf]$ are the smooth approximations of the $\G_k(L,\F)$ coefficients.  

Note that the QFT EA coefficients $\G_{M,k}[g,\vf]$ will not be the same as the usual perturbative QFT coefficients $\G_{K,k}[g,\vf]$, where $\hbar K$ is the momentum cutoff determined by the average edge length in $T(U)$. This is because the coefficents $\G_{K,k}$ are defined on the manifold $U$ where the boundary metrics are flat and the boundary fields are vanishing, see the appendix C.

One can then write
\be \G_{M,k}[g,\vf] = \G_{K,k}[g,\vf] + \D\G_{M,k} [g,\vf]\,,\label{wfuc}\ee
and the corrections can be calculated by using the perturbative expansions of $\G(L,\F)$ and $\G_U(L,\F)$, where $\G_U$ satisfies the equation (\ref{gmea}) for $M = U$, see the appendix D. 

We expect that the corrections $\D\G_k$ will be small compared to $\G_{K,k}$ when 
\be N_U \gg N_0\,,\quad \bar L_0 \approx \bar L_U \,,\ee
where $N_U$ is the number of edges in $T(U)$, $N_0$ is the number of edges in $T(M_0)$, $\bar L_0$ and $\bar L_U$ are the average edge lengths in $T(U)$ and $T(M_0)$, see the appendix D.

The effective (quantum) field configurations $g_{\m\n}( x)$ and $\vf_\a (x)$, $\a = 1,2,...,c$, will be then given as the solutions of 
 the equations of motion
\be{\d\G_{M} \over\d g_{\m\n}( x)} = 0\,, \quad  {\d\G_{M} \over\d\vf_\a (x)} = 0\,,\quad x = (\vec x,\t) \in \S\times[t_i,t_f] \,.\label{qgfc}\ee

In the perturbative QFT regime, we wil have the expansion (\ref{smpe}), so that
\be g_{\m\n} (x) = g_{\m\n}^{(0)}(x) + \hbar g_{\m\n}^{(1)} (x) + \cdots \,,\quad \vf_\a (x) = \vf_{\a}^{(0)} (x) + \hbar \vf_{\a}^{(1)} (x) + \cdots \,,\ee
where $g^{(0)}(x)$ and $\vf^{(0)}(x)$ are solutions of the classical EOM on the manifold $U$. Hence there is a clear relationship between the quantum and the classical field configurations when $\hbar\to 0$, i.e. when the quantum corrections are small.

As far as the Schrodinger wavefunctional is concerned, we have the formula (\ref{qgte}) for $\Psi(q,t)$, so that in the smooth-manifold approximation we will have
\be \Psi(q,t) \approx \Psi_{M} [h(\vec x),\vf(\vec x) , t] \,, \label{swf}\ee
where $t= t_f$, while $h(\vec x)$ is a smooth metric on $\S = \pa M$ that approximates the PL metric $h^{(\t)} (l)$ on $T(\S)$, where $\t$ is a tetrahedron in $T(\S)$. The smooth matter fields $\vf(\vec x)$ approximate the PL fields $\vf(v)$, where $v$ is a vertex in $T(\S)$. Then the probability distribution for a field configuration on $U$ is given by
\be|\Psi(q,t)|^2 = |\Psi(l,\vf,t)|^2 \approx |\Psi_{M} [h(\vec x),\vf(\vec x),t]|^2 \,.\ee

Note that in a QFT, beside the field configurations (for example, the electro-magnetic waves) one can also have the particle trajectories (for example, the elementary particle tracks in the high-energy collision experiments).  Then a simple and natural way to obtain a particle trajectoriy from a field configuration $\vf^{(s)}_\a (\vec x,t)$ ($\a = 1,2,...,c_s$) that corresponds to a field of spin $s$, would be to associate a particle position $\vec x_s (t)$ with a local maximum of 
\be |\vf^{(s)} (\vec x,t)|^2 = \sum_{\a = 1}^{c_s} |\vf^{(s)}_\a (\vec x,t)|^2 \,,\ee
for a fixed value of $t$.

For the sake of simplicity, let us consider the $s=0$ case. By using the QFT effective action $\G_M [g,\vf]$, we can solve the EA equations of motion (\ref{qftem}) with the boundary conditions
\be \lim_{t\to -\infty} \vf(\vec x,t) = \vf_{in} (\vec x )\,,\quad \lim_{t\to +\infty} \vf(\vec x,t) = \vf_{out} (\vec x )\,, \ee
where $\vf_{in}$ and $\vf_{out}$ are given as products of coherent states
\be \vf_{in} = \prod_{m=1}^{n_i} e^{-(\vec x - \vec x_m)^2/(\D l)^2}\,e^{i\vec k_m \cdot\, \vec x}\,, \quad  \vf_{out} = \prod_{n=1}^{n_f} e^{-(\vec x - \vec y_n)^2/(\D l)^2}\,e^{i\vec q_n \cdot\, \vec x}\,.\label{pasc}\ee
Hence $\vf_{in}$ represents the wavefunction of $n_i$ incoming particles with momenta $\vec k_m$ and positions $\vec x_m$ while $\vf_{out}$ represents the wavefunction of $n_f$ outgoing particles with momenta $\vec q_n$ and positions $\vec y_n$. $\D l$ is a characteristic length for this QFT and the vacuum is caracterized by $|\vf| \approx (\D l)^{-3/2}$ for $r \le \D l$. Then we can define the corresponding particle trajectories by determining the set of local maxima of $|\vf(\vec x,t)|^2$ for every $t\in (-\infty,+\infty)$.

In the case of the QFT bound states, e.g. hadrons in QCD, the situation is more complicated. In the case of a mezon, one can try to define the particle trajectories of a quark and the antiquark via the following class of solutions of the EA equations of motion (\ref{qgfc})
\be \psi_\a^a (\vec x,t) \approx g_\a^a (t)\,\exp\left(- {(\vec x - \vec f (t))^2 \over (\D l)^2} \right)\,,\label{qtm}\ee
where $g_\a^a (t)$ and $\vec f (t)$ are to be determined ($\a$ is a spinorial index and $a$ is a color index). Hence  $\max|\psi_\a^a| \approx |g_\a^a (t)|$ for $\vec x \approx \vec f (t)$, and we take $\D l$ to be an appropriate length scale. Therefore the quantum trajectories of a quark and the antiquark will be given by $\vec f(t)$ and $-\vec f(t)$, in the center of mass coordinates.

In the case of three quarks (proton or a neutron), we will look for solutions of the type
\be \psi_\a^a (\vec x,t) \approx g_\a^a (t)\,\exp\left(- \sum_{k=1}^3{(\vec x - \vec f_k (t))^2 \over (\D l)^2} \right)\,,\label{3qt}\ee
so that $\vec f_k(t)$ will be the corresponding quantum trajectories.

\section{Fermion effective action and the WFU}

In the case of fermions there is a slight technical difficulty regarding the definition of the effective action and the WFU which is related to the definition of the fermionic path integral. Namely, the fermionic PI is usually defined as an integral over a Grassmann algebra, so that that the fermionic fields take values in a Grassman algebra\footnote{One can also define the fermionic path integral as an integral over the c-number functions, but in that case the integration variables are not independent because there are second-class constraints for the canonical variables, see \cite{S}. One can then show that the integration of those commuting constrained variables is equivalent to the integration over a set of unconstrained anti-commuting variables.}. Consequently the effective action and the WFU are functionals of Grassmann fields, while we need the functionals of c-number fields. This requires a prescription of how to pass from a Grassmann algebra function to a c-number function.

Let us consider the electrons, so that we have the fermionic field $\psi_\a ( x)$, which is a complex number and a Dirac spinor ($\a=1,2,3,4$), so that we need a definition of the corresponding wavefunctional $\Psi[\psi_\a (\vec x), t]$. Note that in the path-integral formalism one uses the Grassman algebra elements $\chi_\a (x)$ and $\bar\chi^\a (x)$, via the map 
\be \left(\psi_\a (x), \bar\psi^\a ( x)\right) \to \left(\chi_\a (x),\bar\chi^\a ( x)\right)\,,\ee 
where $\bar\psi^\a = (\g_0)^{\b\a} \psi_\b^*$, while $\chi_\a$ and $\bar\chi^\a$ anticommute, i.e.
\be [\chi_\a,\chi_\b]_+ = 0\,,\quad [\chi_\a,\bar\chi_\b]_+ = 0\,, \quad [\bar\chi_\a,\bar\chi_\b]_+ = 0 \,.\ee

Note that $\bar\chi^\a$ and $\chi_\a$ do not have to be related via the same relation as $\psi_\a$ and $\bar\psi^\a$, which would be  $\bar\chi^\a = (\g_0)^{\b\a} \chi^*_\b$, because there is no a natural way to define a complex conjugation in a Grassmann algebra\footnote{One can complexify a Grassmann algebra $\cg$ by constructing $\cg_C = \cg + i\tilde\cg$, where $\tilde\cg = \cg$, so that
$\chi_\a = \th_\a + i \tilde\th_\a$ and $\chi^*_\a = \th_\a - i \tilde\th_\a$.
Then one can have $\bar\chi^\a = (\g_0)^{\b\a} \chi^*_\b$.}. Therefore we will consider $\chi$ and $\bar\chi$ as independent Grassman variables.

The fermionic generating functional is then given by
\be Z_f[\th, \bar\th] = \int \cd\bar\chi \cd\chi \exp\left(\frac{i}{\hbar}\int_M d^4 x \sqrt{g}(\cl_f(\bar\chi,\chi) + \bar\chi\th + \bar\th\chi) \right)\,,\ee
where $\th(x)$ and $\bar\th(x)$ are independent Grassman algebra-valued fields on $M$ and $\bar\chi\th = \bar\chi^\a \th_\a$. Here we assume that all the path integrals are defined via the corresponding PFQG path integrals, see \cite{M4}.

In order to obtain the EA functional $\G_f[\psi(x)]$, we need the generating functional which depends on a c-number function $\psi(x)$. In order to do this, note that a Grassman algebra function  is a vector
\be F(\th,\bar\th) = F_0 + \bar F_1^{k\a}\th_{k\a} + F^k_{1\a} \bar\th^\a_k + F_2^{kl,\a\b} \,\th_{k\a} \th_{l\b}  + \cdots + F_{8n} \,\prod_{k =1}^n \prod_{\a=1}^4 \bar\th^\a_k \, \th_{k\a} \,,\label{gaf}\ee
in the vector space of dimension $2^{8n}$, whose basis is given by the GA elements generated by the products of the Grassmann coordinates $\th_{k\a}$ and $\bar\th^\a_k$, $k=1,2,...,n$. Here $n$ can be considered as the number of vertices in $T(M)$ and $F_l$, $l=0,1,2,...,8n$, are complex numbers. 

We can then define a complex-number polynomial function via the map 
\be (\th_{k\a},\bar\th^\a_k)\to (z_{k\a},\bar z^\a_k)\in(\C^{4n},\C^{4n})\,,\ee 
where now $\bar z_k^\a = (\g_0)^{\b\a} z^*_{k\b}$, so that
\be F(z) =  F_0 + \bar F_1^{k\a} z_{k\a} + F_{1\a}^k \bar z^\a_k + F_2^{kl,\a\b} \, z_{k\a} \,z_{l\b} + \cdots + F_{8n}\, \prod_{k=1}^n\prod_{\a=1}^4\bar z^\a_k  \, z_{k\a} \,.\label{cnf}\ee

Therefore, given a Grassman algebra function $Z_f[\th(x),\bar\th(x)]$ via the equation (\ref{gaf}), we can obtain the function $Z_f [j(x)]$ via the formula (\ref{cnf}),
where $j(x)$ is the fermionic c-number current. Then we can construct the effective action via the Legandre transform
\be \G_f [\psi] = W_f [j] - \int_M d^4x \,(\bar\psi j   + \bar j \psi) \,,\ee
where $W_f = -i\hbar\log Z_f$ and
\be \bar\psi(x) = {\d W_f\over \d j(x)} \,,\quad \psi(x) = {\d W_f\over \d \bar j(x)}\,.\ee

One can also construct a perturbative expansion of $\G_f$ by using
 $$\G_f [\psi(x)] = \int_M d^4x \int_M d^4 y \, \bar\psi(x) \G(x,y)\psi(y) + \cdots $$
 \be = \sum_{m\ge 1,n\ge 1} \prod_{k=1}^m \int_M d^4x_k \prod_{l=1}^n \int_M d^4 y_l \,\bar\psi(x_1)\cdots\bar\psi(x_m) \G_{m,n}(X,Y)\psi(y_1)\cdots\psi(y_n) \,, 
\ee
where $\G_{m,n} (X,Y) = \G_{m,n} (x_1, \cdots, x_n, y_1,\cdots, y_m)$ are the fermionic one-particle irreducible Green's functions. We have supressed the spinor indices, and $\G_{m,n} (X,Y)$ can be obtained from the connected Green's functions
\be G_{m,n} (x_1, \cdots, x_m, y_1,\cdots, y_n) = {\d^{n+m} W_f [\th(x),\bar\th(y)]\over \d\th(x_1)\cdots \d\th(x_m)\d\bar\th(y_1)\cdots\d\bar\th(y_n)}{\Big |_0}\,,\ee
where $W_f[\th,\bar\th] = \log Z_f[\th,\bar\th]$, $F(\th,\bar\th)|_0 = F_0$ and 
\be \log F( \th,\bar\th) \equiv \log F_0 + \sum_{k\ge 1} \frac{F_0^{-k}}{k}\left( \bar F_1^{k\a}\th_{k\a} + F^k_{1\a} \bar\th^\a_k + F_2^{kl,\a\b} \,\th_{k\a} \th_{l\b}  + \cdots \right)^k \,. \ee

In order to define the fermionic WFU functional $\Psi[\psi(\vec x)]$, we will use an analog of the functional $Z_f[\th(x),\bar\th(x)]$ for $M$ with a single boundary $\S$. Let
\be z_f[\th(\vec x),\bar\th(\vec x)] = \int \cd\bar\chi \,\cd\chi\, e^{\frac{i}{\hbar}\left[\int_M d^4 x \sqrt{g}\cl_f(\bar\chi,\chi)
 + \int_\S d^3x \sqrt{h}\left(\bar\chi(\vec x)\th(\vec x) + \bar\th(\vec x) \chi(\vec x)\right)\right]}\,.\ee

Then by using the formulas (\ref{gaf}) and (\ref{cnf}) we obtain the functional $\Psi_f[\psi(\vec x)]$. Hence the probability distribution functional can be calculated as
\be \r_f [\psi(\vec x)]  = |\Psi_f [\psi(\vec x)] |^2 \,.\ee

Note that when considering the Schrodinger equation for a wavefunctional of a fermionic field $\psi(\vec x)$, we need to use the functional of Grassman fields $\chi (\vec x)$ and $\bar\chi(\vec x)$, because a Schrodinger representation of the algebra of the fermionic canonical variables 
\be [\widehat{\bar\psi^\a} (\vec x),\widehat{\bar\psi^\b} (\vec y)]_+ = 0\,,\,  [\widehat{\bar\psi^\a}(\vec x),\widehat{\psi}_\b (\vec y)]_+ = i\hbar\d^\a_\b\, \d(\vec x - \vec y) \,,\, [\widehat{\psi}_\a (\vec x),\widehat{\psi}_\b (\vec y)]_+ = 0\,,\ee
can be realized only through the Grassman variables $\bar\chi(\vec x)$ and $\chi(\vec x)$ as the operators
\be \widehat{\bar\psi^\a} (\vec x) = i\hbar{\d\over\d\chi_\a(\vec x)} + \bar \chi^\a (\vec x) \,,\quad \widehat{\psi}_\a (\vec x) = -i\hbar{\d\over\d\bar\chi^\a(\vec x)} +  \chi_\a (\vec x) \,\,,\label{fcar}\ee
acting on the wavefunctionals
\be\tilde\Psi[\chi(\vec x),\bar\chi(\vec x)] = \sum_{m\ge 1,n\ge 1} \prod_{k=1}^m \int_\S d^3 x_k  \prod_{l=1}^n\int_\S d^3 y_l \,\bar\chi(\vec x_1)\cdots\bar\chi(\vec x_m)
 \g_{m,n} (X,Y) 
\chi(\vec y_1)\cdots\chi(\vec y_n) \,, \label{gwfu}\ee
where $\g_{m,n}(X,Y)$ are tensorial functions, taking values in $\C$. 

Given a Grasman functional $\tilde\Psi[\chi(\vec x),\bar\chi(\vec x),t]$ which is a solution of the Schrodinger equation
\be i\hbar{\pa\tilde\Psi\over\pa t} = \widehat H_f \left(\widehat{\bar\psi}, \widehat{\psi} \right) \tilde\Psi \,,\ee
so that the coefficients $\g_{m,n}(X,Y,t)$ in the expansion (\ref{gwfu}) are known, the corresponding wavefunctional $\Psi_f [\psi(\vec x),t]$ will be given by
\be\Psi_f [\psi(\vec x),t] =\sum_{m,n} \prod_{k,l} \int_\S d^3 x_k  \int_\S d^3 y_l \,\bar\psi(\vec x_1)\cdots\bar\psi(\vec x_m)
 \g_{m,n} (X,Y,t) \psi(\vec y_1)\cdots\psi(\vec y_n)  \,.\label{fcnwf} \ee

When we have bosonic and fermionic fields $\vf(x)$ and $\psi(x)$ such that the classical action has the form
\be S[\vf,\psi] = S_b [\vf] + S_f[\psi,\vf]\,,\ee
then the generating functional is given by
\be Z[j_b ,j_f ] = \int \cd \vf \,e^{i(S_b [\vf]  + \int_M j_b\vf d^4 x)/\hbar}\,Z_f[j_f,\vf]\,. \ee

The WFU is then given by
\be \Psi[\vf,\psi] = \int \cd \tilde\vf \, e^{iS_b [\tilde\vf]/\hbar}\,\Psi_f[\psi,\tilde\vf]\,,\ee
where now $\pa M = \S$.

\section{Conclusions}

We showed that the problems of Bohmian mechanics, which are the violation of the Heisenberg uncertainity relations, absence of quasi-classical trajectories for bound states and the difficulties with obtaining a consistent QFT formulation can be resolved by replacing the dBB equations of motion with the effective action equations of motion. In order for this approach to work, one must generalize the standard QFT effective action, which is only defined for the QFT vacuum state, to a definition which associates an effective action for a wavefunction of the Universe. This is not surprising, given the fact that the dBB wavefunction is really a wavefunction of the Universe. The effective action for a given WFU can be constructed within a path-integral formulation of a quantum gravity theory, so that the initial state for the WFU time evolution is taken to be the Hartle-Hawking state. 

We used a path-integral formulation of quantum gravity given by the PFQG theory, since the PFQG path integral is finite and produces the correct semi-classical expansion of the effective action. Consequently one can calculate the WFU and the corresponding effective action while the EA equations of motion generate the quantum trajectories which can be related to the classical trajectories.

The problem of determination of the time evolution of a field configuration in a QFT is then solved by using the smooth-manifold approximation of the PFQG effective action. This QFT effective action can be approximated for a certain class of triangulations by the usual QFT effective action for GR coupled to matter, where the QFT cutoff is determined by the average edge length in the temporal part of the spacetime, which is given by the manifold $U= \S\times[t_i,t_f]$. The correction to the standard QFT effective action due to a non-trivial WFU can be determined by using the perturbative expansion of the effective action for the manifold $M_0\sqcup U$, see eq. (\ref{wfuc}) and related equations (D.1) and (D.2).

Given a quantum corrected field configuration in the spacetime $U$, one can define the trajectories of the corresponding elementary particles by the local maxima of the modulus square of the field configuration on a spatial section $\S$ of $U$. In this way one obtains a simple and a natural way to describe a transition from having $n_1$ particles on $\S$ at a moment $t_1$ to having $n_2 \ne n_1$ particles at a later moment $t_2$.

Note that this definition of the particle trajectories for a given field configuration in the spacetime can be applied to the fermionic fields, provided that we we can construct the effective action which is a functional of the c-number fermionic fields. This construction is explained in section 6, and it is necessary because in the path-integral quantization the fermionic fields take values in a Grassman algebra. 

In order to obtain our results it was necessarry that the spacetime manifold $M$ has a special topology, given by $M_0\sqcup U$, where $U=\S\times [t_i, t_f]$. Then the inital WFU is given by the Hartle-Hawking wavefuction for the vacuum manifold $M_0$, while the time evolution is determined by the path-integral for the spacetime manifold $M_0\sqcup(\S\times[t_i,t])$, where $t_i < t \le t_f$. In the PFQG theory, the smooth manifold $M$ is replaced by a PL manifold $T(M)$, such that the triangulation of the manifold $U$ is a temporal triangulation. Then the time variable is given by the formula (\ref{tv}). Furthermore, when the number of the edges of $T(M)$ is large and the edge lengths are sufficiently small, the PL metric and the matter PL fields can be approximated by the smooth fields on $M$. In this case the PFQG effective action can be approximated by the effective action $\G_M$ for the corresponding QFT on $M$, which has a cutoff determined by the average edge length in $T(M)$. 

The effective action on the PL manifold $T(M)$ can be easilly calculated perturbatively, at least for the low orders of $\hbar$, see the appendix D,  so that one can calculate the corresponding smooth approximation $\G_M$. The usual QFT effective action is a smooth approximation of the effective action defined on the PL manifold $T(U)$, which we have denoted as $\G_U$. However, the effective action $\G_U$ does not include the contribution from the non-trival initial HH wavefunction. Since the effective action on $T(M)$ contains the HH contribution, we showed in the appendix D that the difference $\G_M - \G_U$ will be small for $N_U\gg N_0$ and $\bar L_U \approx \bar L_0$, where $N_0$, $N_U$, $\bar L_0$ and $\bar L_U$ are the numbers of edges and the average edge lengths in $T(M_0)$ and $T(U)$, respectively. Therefore in this case the usal QFT effective action $\G_U$  will be a good approximation for the effective action of the universe. 

The existence of the smooth-manifold approximation in the PLQG theory solves the problem of defining the field configurations for a QFT in a dBB framework, 
since the QFT wavefunctional can be defined
via the equation (\ref{swf}). In section 6 we showed how to define the WFU for the c-number fermionic fields, since the fermionic path integral is usually defined by using a Grassman algebra valued fields. This problem also appears in the Schrodinger representation of a fermionic QFT, where one has to use the Grassman variables, see eq. (\ref{fcar}). Then given a functional of Grassman fields (\ref{gwfu}), one can obtain a c-number fermionic wavefunctional (\ref{fcnwf}), so that one can calculate the probability density for a fermionic field configuration.

Given the WFU (\ref{swf}), one could still introduce the field theory dBB equations of motion by using the dBB EOM (\ref{dBBft}). However, it is difficult to work in the Schrodinger representation of a QFT, and one can instead use the EA equations of motion (\ref{qftem}), which can be derived by using the standard QFT. Furthermore, one insures the validity of the Heisenberg uncertainity relations, since in the EA case one can use the smooth approximation of the distribution (\ref{hurd}) and one also avoids the problem of the classical limit for the bound-state trajectories in the dBB case. This problem of the dBB EOM was demonstrated in the case of the Hydrogen atom bound states with non-zero angular momentum, see section 2.

Note that the standard dBB argument for the appearence of classical trajectories is that a quantum trajectory obeys the Bohm equation, which is the second Newton law equation with a quantum potential $U(q) -\hbar^2 \nabla^2 R/2mR$, where $\Psi = R\exp(iS/\hbar)$. However, taking the limit $\hbar\to 0$ on a dBB trajectory does not give a classical trajectory, although in this case the Bohm equation becomes the Newton equation for the classical potential. The reason is that the Bohm equation is derived from the first-order dBB equation of motion $m\dot q = \pa S/\pa q$, so that the value of the initial momentum cannot be an arbitrary value. In the EA case, the fundamental equations of motion are of the second, or higher, order in time derivatives, while the classical equations of motion are of the second order in time derivatives, so that the initial values of the coordinates and the momenta are independent. This also implies that one can use the phase-space probability distribution (\ref{hurd}), which guarantees the validity of the Heisenberg uncertainity relations.

In the case of long-range interactions (electro-magnetic forces and gravity), one can have a macroscopic bound state system, like the Sun and the Earth, and we showed that in this case the dBB trajectories will not be close to the classical trajectories. If one adopts the EA equations of motion, then the appearence of the classical trajectories is guaranteed in the classical limit. 

As we have already mentioned, the EA approach allows one to recover the particle trajectories in a QFT. One can look for the solutions of the EA equations of motion that obey the particle scattering boundary conditions (\ref{pasc}), or look for the solutions that resemble the bound states, see eq. (\ref{qtm}) for the 2-particle case or eq. (\ref{3qt}) for the case of three particles. 

The time evolution of the WFU is not expected to be unitary for an arbitrary PL manifold $T(M)$, but only when the triangulation of the manifold $\S\times[t_i,t]$ is a Hamiltonian triangulation. In this case we expect that the corresponding evolution operator $\hat U_T(t)$ to be unitary. The issue of whether $\hat U_T(t)$ is unitary or not, does not affect the problem of the wavefunction collapse for a small subsytem, nor affects the unitarity of a small subsystem time evolution between the measurements, since these phenomena depend on the Hamiltonian of the subsystem and on the nature of the interaction of the subsystem with the rest of the Universe. We expect that unitarity or non-unitarity of the WFU time evolution willl be relevant for the problem of black hole evaporation and  for the problem of particle creation in the Universe.

\section*{Acknowledgements}
Work supported by the FCT project GFM/2025.

\appendix
\section{QM and dBB expectation values}

Without a loss of generality, one can consider a one-dimensional system $(p,q)$. Then
$$ \langle\Psi|\hat p^2 |\Psi\rangle = \int_\R dq\, \Psi^* \left(-\hbar^2 {\pa^2 \Psi\over\pa q^2}\right)= \int_\R dq\, Re^{-iS/\hbar} (-\hbar^2) \left(Re^{iS/\hbar}\right)'' \qquad\qquad  $$
$$\,\, = (-\hbar^2) \int_\R dq \left(RR'' + \frac{i}{\hbar}(2RR'S' + R^2 S'') - \frac{1}{\hbar^2} R^2(S')^2 \right)$$
$$\qquad = \int_\R dq\, R^2 (S')^2 - \hbar^2 \int_\R dq\, RR'' - i\hbar\int_\R dq\,(2RR'S' + R^2 S'') \,.$$

On the other hand
$$\langle p^2 \rangle_\r  = \int_\R dq \int_\R dp \,p^2\, \r (p,q,t) = \int_\R dq\, R^2 (S')^2 \,,$$
where $\r$ is the dBB distribution (\ref{pqbd}) and $f'\equiv {\pa f \over \pa q}$. Therefore
$$ \langle \hat p^2 \rangle \ne \langle p^2 \rangle_\r \,.\qquad (A.1)$$

In the case of a stationary bound state, we have $S' = 0$, so that $\langle p^2 \rangle_\r = 0$, while
$$ \langle \hat p^2 \rangle =- \hbar^2 \int_\R dq\, RR'' = \hbar^2 \int_\R dq\,(R')^2 > 0 \,. $$

\section{Electron trajectory in a Hydrogen atom}

Consider a stationary bound state from the Hydrogen atom energy spectrum given by
$$ \Psi_{n,l,m}(r,\th,\f,t) = R_n(r) P_l(\cos\th)e^{im\f - i\o_n t} \,,\quad m\ne 0\,,$$
where $\o_n = E_n/\hbar$ and $n\ge 2$. The dBB EOM are then given by
$$ m_e\dot x = m\hbar{\pa\f\over \pa x} \,, \quad m_e\dot y = m\hbar{\pa\f\over \pa y} \,, \quad m_e\dot z = m\hbar{\pa\f\over \pa z} = 0 \,,$$
where $\f = \arctan (y/x)$ and $m_e$ is the reduced mass of the electron. We can then choose $z = 0$ so that $x=r\cos\f$ and $y=r\sin\f$ so that
$$ \dot x = \dot r \cos\f - r\dot\f \sin\f = -\m{\sin\f \over r} \,,\quad  \dot y = \dot r \sin\f + r\dot\f \cos\f = \m{\cos\f\over r} \,,$$
where $\m = m\hbar/m_e$. By multiplying the first equation with $\cos\f$ and the second equation with $\sin\f$, and summing them, we obtain
$\dot r =0$, which then gives $r^2\dot\f = \m$, so that
$$ \f = \f_0 + \o t \,,\quad \o = {m\hbar\over m_e r^2} \,.\qquad (B.1)$$

\section{QG path integral and QFT effective action}

The relationship between the usual QM formalism and the PI formalism is given by the Feynman formula
$$ \langle q_2 |\hat U(t_2,t_1) |q_1\rangle = \int \cd q \exp\left(\frac{i}{\hbar}\int_{t_1}^{t_2} L(q,\dot q)\,dt\right)\,, $$
where $q(t_k) = q_k$, $k=1,2$.

In a QFT we use an expectation value of the type
$$\langle \hat q(t) \rangle_{1,2} = \int \cd q \,q(t) \exp\left(\frac{i}{\hbar}\int_{t_1}^{t_2} L(q,\dot q)\,dt'\right)\,,$$
which comes from the generating functional
$$ Z_{1,2}[J(t)] = \int \cd q \exp\left(\frac{i}{\hbar}\int_{t_1}^{t_2} [L(q,\dot q) + J(t)q(t)]\,dt\right)\,,$$
so that
$$ \langle \hat q(t) \rangle_{1,2} = {\d Z_{1,2}\over \d J(t)}\Bv_{J=0}\,. $$

We can then write
$$\langle \hat q(t) \rangle_{1,2}  = \int \cd q \exp\left(\frac{i}{\hbar}\int_{t_1}^{t} L(q,\dot q)\,dt'\right) q(t) \exp\left(\frac{i}{\hbar}\int_{t}^{t_2} L(q,\dot q)\,dt'\right)\,,$$
so that
$$\langle \hat q(t) \rangle_{1,2}  = \langle q_2 |\hat U(t_2,t)\, \hat q \,\hat U(t,t_1) |q_1\rangle \,. $$

Note that when $\hat U(t',t) = \exp(-i\hat H(t'-t)/\hbar)$, then
$$\langle \hat q(t) \rangle_{1,2}  = \langle q_2 |e^{-i\hat H t_2/\hbar}\, \hat q_H(t) \,e^{i\hat H t_1/\hbar} |q_1\rangle \,, $$
where $\hat q_H (t) = \exp(i\hat H t/\hbar)\, \hat q \exp(-i\hat H t/\hbar)$ is the Heisenberg picture operator.

We can then define
$$\langle \hat q(t) \rangle_{\Psi_0} = \langle \Psi_0 |\hat U(t_2,t)\, \hat q \,\hat U(t,t_1) |\Psi_0\rangle \,,$$
which is a solution of the QFT EA equations of motion, when $\Psi_0$ is the vacuum state and  $t_1\to -\infty$ and $t_2 \to +\infty$. 

Note that
$$\langle \hat q(t) \rangle_{\Psi_0}  = \int d^n q_1\int d^n q_2 \Psi_0^* (q_2)\langle \hat q(t) \rangle_{1,2}   \Psi_0(q_1)\,,\quad\quad (C.1)$$
and in the context of a quantum gravity theory, the expression (C.1) can be interpreted as the path integral on the manifold $M_0\sqcup (\S\times[t_1,t_2] )\sqcup M_0$. The relationship between the PFQG effective action and the usual QFT effective action can be understood from the relationships between the generating functions on the different components of the PL manifold $T(M_0\sqcup (\S\times[t_1,t_2] )\sqcup M_0)$. 

Let
$M = M_- \sqcup U$ and $\bar M = M_- \sqcup U \sqcup M_+$ where $U\equiv\S\times[t_i,t_f]$, while $M_\pm$ indicates the manifold $M_0$ with the boundary $\S$ at the time $t_f$ and at the time $t_i$, respectively. We can then write 
$$ Z_{\bar M} = \int d^n q_- \int d^n q_+ \, Z_0( q_-)\, Z_{U}(q_-, q_+) \, Z_0^* (q_+) \,,$$
and
$$ Z_{\bar M} (\bar J) = \int d^n q_- \int d^n q_+ \, Z_0(J_-, j_-, q_-)\, Z_{U}(q_-, \tilde J, q_+) \, Z_0^* (q_+,j_+,  J_+) \,,$$
where
$$ \bar J = (J_-, j_-, \tilde J, j_+, J_+) \,, \quad Q = (Q_-, q_-, \tilde Q, q_+, Q_+) \,, $$
$Q_k = (L_k, \F_k)$, $q_\pm = (l_\pm,\vf_\pm)$ and $n=n_l + n_\vf$, where $n_l$ is the number of edges and $n_\vf$ is the number of vertices of $T(\S)$.  The simbol $Z_0^*$ means that we take $e^{-iS/\hbar}$ instead of $e^{iS/\hbar}$ in the integrand, where $S$ is the classical action on $T(M_+)$. 

In the standard QFT we are not interested in the dynamics of $Q_\pm$, so that we use
$$\tilde Z_{U} (\tilde J) = Z_{\bar M} (0,0,\tilde J,0,0) = \int d^n q_- \int d^n q_+ \, Z_0( q_-)\, Z_{U}(q_-, \tilde J, q_+) \, Z_0^* (q_+) \,,$$
$$ = \int d^n q_- \int d^n q_+ \, \Psi_0( q_-)\, Z_{U}(q_-, \tilde J, q_+) \, \Psi_0^* (q_+) \,.$$
Consequently
$$ \tilde\G_U(\tilde Q) = \tilde W_U (\tilde J)  - \tilde J \tilde Q \,,$$
where now $\tilde Q = \pa \tilde W_U /\pa \tilde J$ and $\tilde W_U = -i\hbar\,\log \tilde Z_U$.

Note that one can also use 
$$ \G_{U} (q_-, \tilde Q,q_+) = W_U (j_-, \tilde J,j_+)   - j_-q_-  - \tilde J \tilde Q - j_+q_+   \,,$$
where $W_U = -i\hbar\, \log Z_U$, $\tilde Q = \pa W_U /\pa \tilde J$, $q_\pm = \pa W_U /\pa j_\pm$, $\tilde J \tilde Q = J_L L + i J_\F \F$ and $jq = j_l j + ij_\vf \vf$. 

When
$$ \Psi_0(q) = \d(q-q_0) = \d(l-l_0)\d(\vf - \vf_0) \,,\quad\quad (C.2) $$
where $l_0$ gives a flat metric on $T(\S)$ and $\vf_0 = 0$ we have
$$\tilde\G_U (L,\F) =  \G_U (q_0, L, \F, q_0) \,. $$

If $N_U\to\infty$ such that $L_\e = O(1/N_U)$ for all $\e\in T(U)$ and for the trivial WFU (C.2) then
$$\tilde\G(L,\F) \approx \tilde\G_U [g,\vf] \equiv \G_{U,K} [g,\vf] \,,$$
where $\G_{U,K}$ is the usual QFT EA for the momentum cutoff $\hbar K$, which is proportional to $\hbar/\bar L_U$, where $\bar L_U$ the average edge length in $T(U)$. 

\section{The WFU correction}

When the WFU is nontrivial, instead of using the effective action $\tilde\G_U(\tilde Q)$, it is easier to use the effective action for the manifold $M = M_0\sqcup U$, which we denote as $\G_M (Q_-,q_-,\tilde Q, q_+)$. We can then use
$$\G_M(Q_-,q_-,\tilde Q, q_+) = \G_{U}(q_-,\tilde Q, q_+) + \D\G_M(Q_-,q_-,\tilde Q, q_+) \,,$$
and the correction $\D\G_M$ can be calculated perturbatively by using the perturbative expansions 
$$\G_U (q_-, \tilde Q,q_+) = \sum_{k \ge 0}\hbar^k \,\G_{U,k}(q_-, \tilde Q,q_+)\,,$$ 
and
$$\G_M (Q_-,q_-, \tilde Q,q_+) = \sum_{k \ge 0}\hbar^k \,\G_{M,k}(Q_-,q_-, \tilde Q,q_+)\,,$$ 
when $|L_\e| \gg l_P$ and $|\tilde\vf_v| < 1$ for $\e,v \in T(M)$.

We then have
$$\G_{M,0} (Q_M) = S(Q_-,q_-) + S (q_-, \tilde Q , q_+) \equiv S_0 + S_U  \,,$$
and
$$\G_{M,1} (Q_M) = \frac{i}{2} Tr(\log (S_0 + S_U)'') - i \log\m(L_M) \,,\quad\quad\quad (D.1)$$
while
$$\G_{U,0} = S_U\,,\quad \G_{U,1} (Q_U) = \frac{i}{2} Tr(\log S_U'') - i \log\m(L_U) \,,\quad\quad\quad (D.2)$$
where $Q_U = (q_-,\tilde Q,q_+)$ and $S''$, $S''_U$ denote the corresponding Hessian matrices \cite{MVb}. The higher-order corrections $\G_k$  are functions of the higher-order derivatives of $S(L,\F)$ and of the higher-order derivatives of $\log\m(L)$, and a $\G_k$ function is determined by summing the evaluations of the connected graphs with $k$ loops, see \cite{MVb}. Consequently we can write
$$ \G_{M,k} (Q_M) = \G_{U,k}(Q_U) + \D\G_{M,k}(Q_M)\,,$$
for $k=0,1,2,...$ .

Given an arbitrary manifold $M$, then on $T(M)$ we can rewrite the perturbative expansion (\ref{plpe}) as
$$ {\G(L,\F)\over\hbar} = {S_R(L) + \tilde S_m(L,\F)\over l_P^2} \, +\, \G_1(L,\F) \,+\,  l_P^2 {\G_2(L,\F)\over G_N} \,+\,l_P^4 {\G_3(L,\F)\over G_N^2} \,+ \cdots$$
$$ \equiv {\tilde S (L,\F)\over l_P^2} \,+\, \sum_{k\ge 1} l_P^{2(k-1)}\,\tilde \G_k (L,\F) \,. \qquad\qquad\qquad\qquad\quad (D.3)$$

One can show that for large $N$
$$ \tilde S = O( N (\bar L)^2 ) \,,\quad \tilde\G_1 = O(N) \,,\quad\quad\quad\quad\quad (D.4)$$
where $N$ is the number of edges in $T(M)$ and $\bar L$ is the average edge length in $T(M)$.  From the expansion (D.3)  and the result (D.4) we expect to have for $k>1$
$$ \tilde\G_k = O\left( {N\over (\bar L)^{2(k-1)}}\right) \,.$$

Consequently
$$ \tilde S_U = O(N_U \bar L_U^2)\,, \quad \tilde S_0 = O(N_0 \bar L_0^2)\,,$$
so that $|S_U| \gg |S_0|$ for $N_U \gg N_0$ and $\bar L_U \approx \bar L_0$. Similarly,
$$\tilde\G_{U,k} = O\left( {N_U\over (\bar L_U)^{2(k-1)}}\right)\,, \quad \tilde\G_{M,k} =O\left({N_U + N_0 \over (\bar L_M)^{2(k-1)}} \right)\,,$$
so that for $k\ge 1$ we obtain
$$ |\G_{U,k}| \approx |\G_{M,k}|\,, $$
for $N_U \gg N_0$ and $\bar L_U \approx \bar L_M$, which is a consequence of $\bar L_U \approx \bar L_0$. This then implies
$$\G_M(Q_M) \approx \G_U (Q_U)\,,$$
and 
$$ |\G_U(Q_U)| \gg |\D\G_M (Q_M)|\,.$$

\end{document}